\documentclass[epsfig,12pt,onecolumn]{article}
\usepackage{amsfonts}
\usepackage{amssymb}
\usepackage{amsmath}
\usepackage{multicol}
\usepackage{graphicx}
\usepackage{float}
\usepackage{caption}

\input{tcilatex}
\makeatletter \@addtoreset{equation}{section}

\def \be{\begin{equation}}
\def \ee{\end{equation}}
\def \bea{\begin{eqnarray}}
\def \eea{\end{eqnarray}}

\newcommand{\nc}{\newcommand}
\nc{\al}{\alpha} \nc{\bib}{\bibitem} \nc{\la}{\lambda}
\nc{\C}{\mbox{\hspace{1.24mm}\rule{0.2mm}{2.5mm}\hspace{-2.7mm} C}}
\nc{\R}{\mbox{\hspace{.04mm}\rule{0.2mm}{2.8mm}\hspace{-1.5mm} R}}

\begin{document}

\title{Entropy as logarithmic term of the central charge and modified
Friedmann equation in AdS/CFT correspondence}
\author{M. Bousder$^{1}$\thanks{%
mostafa.bousder@um5.ac.ma}, E. Salmani$^{2,3}$, and H. Ez-Zahraouy$^{2,3}$ \\
$^{1}$ {\small Faculty of Science, Mohammed V University in Rabat, Morocco}\\
$^{2}${\small Laboratory of Condensend Matter and Interdisplinary Sciences,
Department of physics,}\\
\ {\small Faculty of Sciences, Mohammed V University in Rabat, Morocco}\\
$^{3}${\small CNRST Labeled Research Unit (URL-CNRST), Morocco}}
\maketitle

\begin{abstract}
This paper is about the extended thermodynamics of AdS black holes and its
relation to CFT thermodynamics. The logarithmic term of the central charge
is interpreted as black hole entropy. We have obtain a modified Friedmann
equation from the Smarr formula. We find that the AdS radius is the critical
shadow radius. We obtain the Hawking-Bekenstein formula with logarithmic
corrections, which depends on the central charge. The real gas in AdS is a
dual of an ideal gas in CFT. This work can be extended to the AdS-Kerr black
holes.

\textbf{Keywords:} AdS-CFT Correspondence, Black Holes, Van der Waals gas.
\end{abstract}

\section{Introduction}

A new class of black hole entropy called Barrow entropy Ref. \cite{PLB1}
describes quantum gravitational effects, which generate a fractal structure
on the surface of the black hole. The influence of the Barrow entropy
effects on the Friedmann equations was explored in Ref. \cite{PRD1} and by
using f(R) gravity Ref. \cite{PLB2}. Particularly, this entropy is applied
with the generalized logarithmic modification and its relationship with
Tsallis entropy Ref. \cite{PLB3}. This approach offers a new perspective to
study the equations of cosmology using this new entropy.\newline
In Ref. \cite{EJPC1} has been exploited to infer the generalized second law
of thermodynamics in a universe filled with matter and dark energy fluids.
Some recent works have been done to study the observational constraints on
Barrow holographic dark energy Ref. \cite{EJPC2}, such as the Big bang
nucleosynthesis constraints Ref. \cite{PLB4} and constraints from M87* and
S2 star observations Ref. \cite{Un}. This entropy sheds light on the
holographic dark energy and thermodynamic aspects of black holes Refs. \cite%
{PLB5, PLB50}.\newline
Recently, in the context of the gauge/gravity duality, M. R. Visser Ref.
\cite{V0} proposed a novel holographic central charge $C=L^{d-2}/G$, which
has attracted great attention. He finds that the Euler equation is dual to
the generalized Smarr formula for black holes in the presence of a negative
cosmological constant. This theory sheds light on the holographic and
thermodynamic aspects of black holes \cite{NPB1}. In comparison, the authors
in Refs. \cite{RE1,RE2} have been proposing a restricted phase space
approach, in which the pressure $P$ and the volume $V$ are absent in the
thermodynamic description for black holes. The AdS/CFT correspondence Ref.
\cite{IN0} describes the asymptotically anti-de Sitter (AdS) black holes by
the dual conformal field theory (CFT) at finite temperature Ref. \cite{PO4}.
Cardy's formula governing the microstates of black holes also leads to a
macroscopic description of area law for Kerr-AdS black holes Ref. \cite{IN1}%
. In d-dimensional CFT there are several candidates for the central charge $C
$. In the dual CFT, the chemical potential $\mu $ is the conjugate
quantities of the central charge, this corresponds to including $\left( \mu
,C\right) $ as a new pair of conjugate thermodynamic variables Ref. \cite%
{IN2}. The Born-Infeld parameter modifies the thermodynamic volume and the
chemical potential Ref. \cite{IN3}. In Ref. \cite{RE7}, a novel
thermodynamic volume was obtained, leading to a new understanding of the Van
der Waals behavior of charged AdS black holes in CFT. To describe the phase
transition, it's useful to study the behavior of specific heat,
compressibility and volume expansivity at the critical point. In this
regard, the cosmological constant and the gravitational Newton's constant
are varied in bulk. In the near horizon of the Kerr black hole with
zero-temperature, the microscopic entropy of the CFT is calculated using the
Cardy formula, which coincides with the extremal case of the macroscopic
Kerr black hole entropy Ref. \cite{IN4}. The time scale of the small-large
black hole phase transition shifts to the left with the increase in
temperature. This corresponds to a first passage process that occurs at a
shorter time for higher temperatures Ref. \cite{RE10}. \newline
Our analysis reveals that the CFT central charge leads to a complete
description of black hole thermodynamics, which represents the equation of
state and the $P-V$ criticality Ref. \cite{PO3}. The thermodynamic variables
of the black holes are obtained in terms of a new parameterization. This
parameter is the ratio between the horizon radius with respect to the length
of d-dimensional AdS to the power of d, which will allow us to describe
several holographic properties and thermodynamic parameters of the black
hole with simple expressions. This approach offers a new perspective on the
central charge from this more general viewpoint.\newline
Next, we set $c=\hbar =1$.\newline
The article is planned as follows: In section 2, we study the CFT central
charge in the black hole. In Section 3, we derive the equation of state PV
and the modified Friedmann equation from the relation between the AdS
pressure and the cosmological constant. In Section 4, we discuss the
evolution of a free photon orbiting around the 4-dimensional black hole. In
section 5, we analyze the thermodynamic stability of the black hole under
the perturbations by using specific heat. In Section 6, we study entropy,
bulk sound speed, and equation of state PV describing the interior of the
ideal gas. We discuss the duality between holographic CFT and AdS
thermodynamics. We show also the Van der Waals equation of the AdS fluid.

In Section 7, we explore an application for AdS-Kerr black hole. We conclude
our findings in Section 8.

\section{AdS/CFT and Barrow entropy}

In this section, we propose a method to find the equation of state $PV$ with
a new parameterization. The Einstein-Hilbert bulk action Ref. \cite{PO3} is
given by%
\begin{equation*}
I_{bulk}=\frac{1}{16\pi G}\int d^{d+1}x\sqrt{-g}\left( R-2\Lambda \right) .
\end{equation*}%
By solving the field equation, we use the black hole metric Ref. \cite{RE1}%
\begin{equation}
ds^{2}=-f(r)dt^{2}+\frac{1}{f(r)}dr^{2}+\cdots +r^{2}d\Omega _{d-2}^{2}.
\label{a1}
\end{equation}%
Generally, the event horizon is located at $f(r)=0$. The asymptotic behavior
($r\rightarrow \infty $) of $f(r)$ is%
\begin{equation}
f_{\infty }(r)\approx \frac{r^{2}}{L^{2}}.  \label{a3}
\end{equation}%
where $L$ is a fixed AdS radius. Let us begin by explaining the importance
of the expression of $f_{\infty }$. If we replace $r$ with the black hole
radius, we find several physical properties like entropy, mass, pressure,
central charge, etc. The boundary metric is obtained via Ref. \cite{RE1}: $%
ds_{CFT}^{2}=\lim_{r\rightarrow \infty }\frac{L_{0}^{2}}{r^{2}}ds^{2}$,
where $L_{0}$ denotes the maximum value of $L$. The thermodynamic variables
for the bulk black hole are equivalent to those of the dual CFT, if only if $%
L=L_{0}$, which is equivalent to the AdS/CFT correspondence Ref. \cite{RE1}.
In this case, the partition function on both sides is $Z_{CFT}=Z_{AdS}=\exp
(-I_{E})$, where $I_{E}$\ is the Euclidean action. The central charge $C$
encodes the level in the CFT is%
\begin{equation}
C=\frac{L^{d-2}}{G}.  \label{a6}
\end{equation}%
Here, $G$ is Newton's constant in $d+1$ dimensional AdS space-time and $C$
varied the $d$ dimensional CFT. A key concept in understanding AdS black
hole thermodynamics is the relation between the thermodynamic parameters and
the central charge. The Gibbs free energy $W$ of the CFT can be defined via $%
W=-T\log Z_{CFT}=\mu C$, where $\mu $ is the chemical potential with
\begin{equation}
\mu =\frac{GM}{2L^{2}}.  \label{a7}
\end{equation}%
Here $\mu \ $is conjugate to $C$. When $\mu $ changes sign to become
positive, there is the presence of quantum effects Ref. \cite{RE4}. The
numbers of microscopic degrees of freedom for the black hole Refs. \cite%
{RE1,RK3} are proportional to its areas and are given by
\begin{equation}
N_{AdS}=\frac{L^{2}}{G}\text{, \ \ }N_{CFT}=\frac{A}{G}.  \label{a8}
\end{equation}%
where $A$ is the area of the unit 3-sphere. The numbers $N_{AdS}$ and\ $%
N_{CFT}$ are the degrees of freedom on the bulk AdS and CFT horizon,
respectively. For simplicity of explanation, we first consider the number $%
N_{CFT}$ of states which are defined through the event horizon, while it is
not the total number of states in CFT because there are black holes with two
horizons. Note that the number $N_{CFT}$ is equivalent to the
Bekenstein-Hawking formula Ref. \cite{BHF1,BHF2} $S_{BH}=\frac{A}{4G}$. For
this we propose that there is an entropy $S_{AdS}$ behind the number $N_{AdS}
$ of the bulk enclosed by the horizon. The corresponding numbers degrees of
freedom and modified central charge $C$, by restructuring Barrow entropy,
analogues to the Bekenstein formula, leads to an effective Barrow entropy
Ref. \cite{PLB1}:%
\begin{equation}
S=\left( \frac{A}{A_{0}}\right) ^{1+\frac{\delta }{2}},  \label{as1}
\end{equation}%
which is given by where $0\leq \delta \leq 1$, $A$ is the black hole horizon
area and $A_{0}$ is the Planck area. When $\delta =0$, the area law is
restored, i.e. $S=\frac{A}{4G}$ $\left( \text{where }A_{0}=4G\right) $.
While $\delta =1$ represents the most intricate and fractal structure of the
horizon. It is important to remark that the equivalence\ between Eq. (\ref%
{a8}) and the Barrow entropy Ref. \cite{PLB1}. From Eq. (\ref{a8}), one can
represent the two entropies as%
\begin{equation}
S_{AdS}=\frac{L^{2}}{4G}\text{, \ \ }S_{BH}=S_{CFT}=\frac{A}{4G}.
\label{as2}
\end{equation}%
Further considering the generalized entropy $S$ reduces to%
\begin{equation}
\mathcal{S}=\left( \frac{\mathcal{A}}{A_{0}}\right) ^{1+\frac{\delta }{2}},
\label{as3}
\end{equation}%
with $A_{AdS}=L^{2}$ and $A_{CFT}=A$. Next, we propose a specific form of
AdS emergence by using $L=H^{-1}$ Refs. \cite{D9,MB0} and Eqs. (\ref{a6},\ref%
{as3}), hence the entropies reduces to%
\begin{equation}
S_{AdS}=\frac{C}{4}H^{d-4}\text{, \ \ }S_{CFT}=\frac{C}{4}AH^{d-2},
\label{as4}
\end{equation}%
where $H$\ is the Hubble parameter. We notice that $S_{AdS}=\frac{C}{4}$\ is
a constant in $d=4$. The apparent horizon radius Ref. \cite{PRD1} is given
by $r_{A}^{2}=\frac{1}{H^{2}+\frac{k}{a^{2}}}$.

\section{Modified Friedmann equation}

Our starting point is the d-dimensional cosmological constant is%
\begin{equation}
\Lambda =-\frac{\left( d-1\right) \left( d-2\right) }{2L^{2}}.  \label{a9}
\end{equation}%
The AdS pressure is associated with a negative cosmological constant $P=-%
\frac{\Lambda }{8\pi G}.$Using the above relations we get $P=\frac{\left(
d-1\right) \left( d-2\right) }{16\pi GL^{2}}$, which reduces to $PL^{d}=%
\frac{\left( d-1\right) \left( d-2\right) }{16\pi }C$. Assuming that the
black hole is contained in a spherical box of radius $L$ centered at $r=0$.
The energy $PV$ required for the d-dimensional volume of a Euclidean ball of
radius $L$ and of radius $r_{H}$, respectively:%
\begin{equation}
V_{L}=\frac{\pi ^{d/2}}{\Gamma \left( \frac{d}{2}+1\right) }L^{d}\text{, \ \
\ }V=\frac{\pi ^{d/2}}{\Gamma \left( \frac{d}{2}+1\right) }r_{H}^{d}.
\label{a10}
\end{equation}%
which leads to
\begin{equation}
PV_{L}=\frac{\pi ^{\frac{d}{2}-1}\left( d-1\right) \left( d-2\right) }{%
16\Gamma \left( \frac{d}{2}+1\right) }C  \label{a11}
\end{equation}%
For $d=4$, we get $PV=\frac{3\pi }{16}C$, which shows that the $PV$ energy
is expressed as a function of the central charges of CFT$_{d}$. Therefore,
one would expect that the $\chi $ is a parameter that makes it possible to
transform the term $PV$ into the Gibbs free energy $W=\mu C$. This result
shows that energy is encoded on the horizon as quantum information. We
likewise identify%
\begin{equation}
\chi =\frac{V}{V_{L}}=\frac{r_{H}^{d}}{L^{d}}.  \label{a12}
\end{equation}%
Similarly we have $\chi =\left( N_{CFT}/\pi N_{AdS}\right) ^{d/2}$. It is
obvious that from dimensional analysis, the term
\begin{equation}
\mu _{d}=\frac{\pi ^{\frac{d}{2}-1}\left( d-1\right) \left( d-2\right) }{%
16\Gamma \left( \frac{d}{2}+1\right) }\propto \Lambda L^{2},
\end{equation}%
is a chemical potential. While, $\mu _{d}$\ is different from $\mu $ Eq. (%
\ref{a7}), because $\mu $ represents the chemical potential of CFT and $\mu
_{d}$ is a geometric quantity that has the dimension of chemical potential.
In section 4, we will see that $\mu _{d}$ plays an important role in the
thermodynamic description of both CFT and AdS. The $\mu _{d}$ depends on
both the cosmological constant and the AdS radius. From Eq. (\ref{a11}) we
get
\begin{equation}
PV=\mu _{d}\chi C.  \label{a14}
\end{equation}%
Therefore, the $PV$ Eq. (\ref{a14}) describes a CFT fluid from the central
charge $C$. Note that for the ideal gas law ($PV=T$) we obtain $T\propto C$.
The energy $PV$ required to displace the vacuum energy of the internal
energy $E$ Ref. \cite{RE5}: $PV=M-E$. For $d=4$, we have$\ V=4\pi r_{+}^{2}/3
$ Ref. \cite{PO6} and if $\mu _{d}=\mu $ we get $PV=\chi \frac{M}{2}L^{d-4}.$
Comparing the resulting law of emergence in the Eq. (\ref{as1}) with that
obtained for Eq. (\ref{a6}), we get $C\equiv \frac{4L^{d-2}}{A_{0}}$. Then,
the parameter $\chi $ can be further rewritten as $\chi =\frac{PV}{\mu _{d}C}
$. We consider only metrics of the form%
\begin{equation}
\chi =\frac{A_{0}PV}{\mu _{d}4L^{d-2}}.
\end{equation}%
\begin{equation}
A_{0}\equiv \frac{4L^{d-2}}{C}.
\end{equation}%
The energy-momentum tensor for perfect fluid is given as%
\begin{equation}
T^{\mu \nu }=\left( \rho +p\right) u^{\mu }u^{\nu }+pg^{\mu \nu },
\label{g1}
\end{equation}%
where $\rho (r)$ is the energy density, $p$ is the pressure of\ the
homogeneously distributed matter and $u^{\mu }$ is the four-velocity of the
fluid. Padmanabhan Ref. \cite{O1} argued that the special expansion of our
Universe can be understood as the consequence of the emergence of space. The
work density associated with this volume change is defined as \cite{PRD1}.
We can reproduce a modified Friedmann equation based on the Barrow entropy
by using the Smarr formula Ref. \cite{CQG1} in d-dimensions:%
\begin{equation}
\frac{d-3}{d-2}M=TS_{AdS}+TS_{CFT}+\mu _{d}C-\frac{2}{d-2}PV.  \label{g2}
\end{equation}%
where $\mu _{d}$ is the chemical potential. Using Eq. (\ref{as4}), the above
equation can be further rewritten as a modified Friedmann equation%
\begin{equation}
H^{d-4}+AH^{d-2}=\left( \frac{M}{V}+\frac{2}{d-3}P\right) \left( \frac{d-3}{%
d-2}\right) \frac{4\mathcal{Z}}{PC}-\frac{4\mu _{d}}{T}.  \label{g3}
\end{equation}%
where $Z=\frac{PV}{T}$ is the compressibility factor. For an ideal fluid,
the compressibility factor becomes $Z=1$ The case of $Z>1$, describes the
repulsion between the fluid particles and generates force in an outward
direction (repulsive). In the case of $Z<1$, there is a strong attraction
between these particles. Eq. (\ref{g3}) looks like both the Van der Waals
equation and the Friedmann equation. This equation is very important because
it connects both a cosmological term which is parametrized by $H$ and a term
of the thermodynamics of the black hole.\ For $d=4$, arrive at $H^{2}+\frac{1%
}{A}+\frac{4\mu _{d}}{T}=\left( \rho +2P\right) \frac{2\mathcal{Z}}{PC}$,
with $\rho =\frac{M}{V}$. Note the presence of the crucial $PV$ term; we
shall demonstrate the need for this term in an example below.

\section{Photon sphere and shadow radius}

Next, we analyze the evolution of a free photon orbiting around the
equatorial orbit of the 4-dimensional black hole along a null geodesic. The
photon Lagrangian for this system can be expressed as Ref. \cite{w3} $L=%
\frac{1}{2}\left[ -f(r)\dot{t}^{2}+\frac{1}{f(r)}\dot{r}^{2}+r^{2}\dot{\phi}%
^{2}\right] $ with $\dot{r}=\partial r/\partial \lambda $ and the affine
parameter $\lambda $. The generalized momenta is $p_{\mu }=\frac{\partial
\mathcal{L}}{\partial \dot{x}^{\mu }}=g_{\mu \nu }\dot{x}^{\nu }$ with $%
p_{r}=\dot{r}/f(r)$.\ There exist two Killing fields $\partial /\partial t$
and $\partial /\partial \varphi $,\ we obtain the energy total energy $%
E=-p_{t}=f(r)\dot{t}^{2}$ and orbital angular momentum $\hat{L}=p_{\varphi
}=r^{2}\dot{\varphi}$ which are constants. We easily get the equation of
radial motion $V_{eff}+\dot{r}^{2}=0$. Since $\dot{r}^{2}>0$, we obtain the
condition: $V_{eff}<0$, i.e. the photon can only survive in the range of
negative effective potential. Thus the effective potential for a free photon
can be expressed as%
\begin{equation}
V_{eff}=\frac{\hat{L}^{2}}{r^{2}}f(r)-E^{2},  \label{a18}
\end{equation}%
Using the above relations we have $V_{eff}=\dot{\varphi}^{2}r^{2}f(r)-\dot{t}%
^{4}f^{2}(r).$ From Eq. (\ref{a3}), we obtain the AdS potential in the
horizon ($r=r_{H}$) as
\begin{equation}
V_{eff}^{\infty }\left( r_{H}\right) \ =\left( L^{2}\dot{\varphi}^{2}-\dot{t}%
^{4}\right) \chi _{d=4}.  \label{a19}
\end{equation}%
where$\chi _{d=4}=\frac{r_{H}^{4}}{L^{4}}$. Using the condition $V_{eff}<0$,
we get $\left( L\dot{\varphi}/\dot{t}^{2}\right) ^{2}<1$. We notice that the
photon follows the geometry characterized by $\chi _{d=4}$ in the horizon,
which relates both the quantum behavior of CFT and the AdS geometry. The set
of photons in the AdS horizon can only survive in the range of the special
metric $\chi _{d=4}$. This generates a gas of energy $PV$ Eq. (\ref{a14}),
which is proportional to the CFT central charge. In the following, we mainly
discuss the relation between the photon sphere radius and the special metric
$\chi $. The circular unstable photon sphere satisfies the condition $\dot{r}%
=\ddot{r}=0$, yielding
\begin{equation}
V_{eff}\left( \chi \right) =\frac{\partial }{\partial \chi }V_{eff}\left(
\chi \right) =0\text{ },\text{\ }\frac{\partial ^{2}}{\partial \chi ^{2}}%
V_{eff}\left( \chi \right) <0.  \label{a20}
\end{equation}%
By solving above conditions: $\partial _{\chi }V_{eff}=0$, the radius of the
photon sphere $r_{ps}$ can be derived as $rf^{\prime }(r_{ps})-2f(r_{ps})=0$
or $rf^{\prime }(\chi _{ps})-2f(\chi _{ps})=0$ with $\chi _{ps}=\frac{%
r_{ps}^{d}}{L^{d}}$. Secondly, solving $V_{eff}=0$ for a spherically
symmetric black hole, we get the shadow radius Ref. \cite{w4}
\begin{equation}
R_{S}=\frac{r_{ps}}{\sqrt{f\left( r_{ps}\right) }}\text{, }R_{S}^{d}=L^{d}%
\frac{\chi _{ps}}{f^{\frac{d}{2}}\left( \chi _{ps}\right) }.  \label{a21}
\end{equation}%
It is immediately clear that there is a relationship between the shadow and
AdS radius. Here, one starts by considering the metric Eq. (\ref{a3}), one
obtains%
\begin{equation}
R_{S}\left( f_{\infty }\right) =L,  \label{a22}
\end{equation}%
i.e. the AdS radius is the critical shadow radius. From Eq. (\ref{a3}), we
have $L=\frac{r}{\sqrt{f_{\infty }(r})}$, i.e. $L$ represents the shadow
radius in the branch of AdS$_{d+1}$, on the other hand, $R_{S}$ is the
shadow at d dimensions. This connection between AdS space-time and black
hole shadow shows the presence of holographic entanglement entropy in the
thermodynamic description of the black hole Refs. \cite{RE6, RT00}.

\section{Entropy and central charge}

\begin{figure}[H]
\centering\includegraphics[width=11cm]{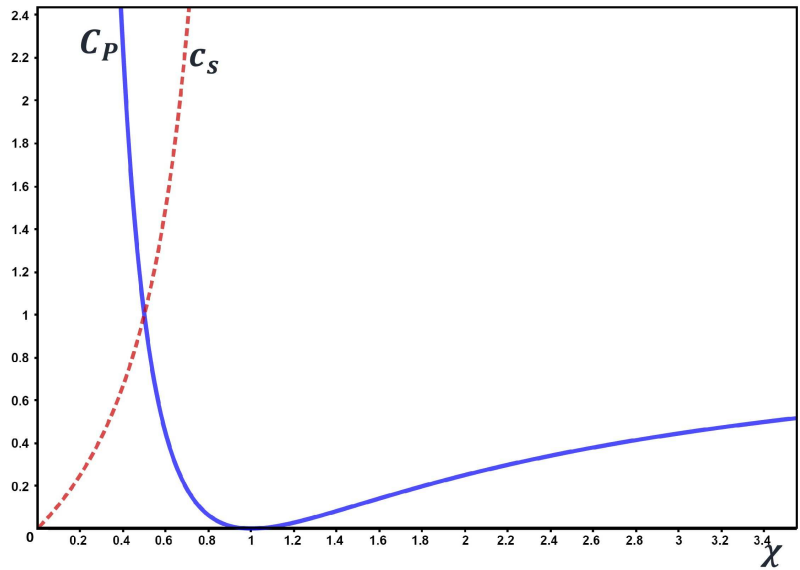}
\caption{Curves of specific heat vs $\protect\chi $\ (solid line) and\ sound
speed vs $\protect\chi $\ (dashed line).}
\label{F1}
\end{figure}

The black hole mass $M$\ is identified with the thermodynamic enthalpy $H$
with $dH=dM$. The black hole first law reads
\begin{equation}
dM=TdS+\mu _{d}dC+VdP+\Phi dQ+\Omega dJ.  \label{b1}
\end{equation}%
This equation mixes the volume and boundary description of a black hole
because $C$ is the central boundary charge and $P$ is the volume pressure.
This equation involves an AdS black hole space-time and a dual CFT Ref. \cite%
{RE1, RE7} of basic structure of Visser's framework Ref. \cite{V0}. The
parameters $V$ and $\mu _{d}$ are the conjugate quantities of the pressure $P
$ and the central charge $C$, respectively.\ Notice that $\mu _{d}$ has
units of the chemical potential. This construction is a combination of the
extended phase space thermodynamics and that of Visser's holographic
thermodynamics Ref. \cite{V0}, as indicated by Eq. (\ref{b1}). While from
Refs.\cite{RE1,RE2}, $P$ and $V$ are absent in the thermodynamic description
for black holes. The key point lies in that only in the restricted phase
space approach there can be a perfect Euler relation Refs.\cite{RE1, RK1},
which is required for the thermodynamic system to be extensive. Extensivity
implies that the intensive quantities like temperature and$\ $chemical
potential are zeroth order homogeneous functions that do not depend on the
size/scale of the system, and this is a pre-requisite for the zeroth law of
thermodynamics to hold. Here, we want to show the influence of the center
charge in the state equation $P-V$. Next, by fixing the bulk pressure $P$, $Q
$ and $J$ Ref. \cite{PO14}, one can obtain the corresponding entropy as%
\begin{equation}
S=\int \frac{dM}{T}-\mu _{d}\int \frac{dC}{T}.  \label{b2}
\end{equation}%
The entropy also depends on the central boundary charge $C$. For an ideal
gas we get the Hawking-Bekenstein formula: $S_{A}=\int \frac{dM}{T}=\pi
r_{+}^{2}/G=A/4G$ and $PV=T$ Eq. (\ref{a14}), which leads to%
\begin{equation}
S=\frac{A}{4G}+\frac{1}{\chi }\ln \frac{C_{0}}{C},  \label{b3}
\end{equation}%
where $C$ is an ultraviolet (UV) cutoff and $1/\chi $ represents the unit
element of entropy. In order to ensure $S>S_{A}$ we need $C>C_{0}$. If $%
C<C_{0}$\ we get $S<S_{A}$. For $C=C_{0}$\ or $\chi \rightarrow \infty $ we
obtain $S=S_{A}$. We also remark that there is the presence of the parameter
$\chi $, which represents the map of information between AdS and CFT. The
macroscopic Bekenstein-Hawking area law assuming Cardy formula $S=\frac{A}{4G%
}=\frac{\pi ^{2}}{3}\left( C_{R}T_{R}+C_{L}T_{L}\right) $. The right/left
central charge in the 2D CFT description for 4D black holes is suggested to
be equal in each sector $C_{R}=C_{L}$. The presence of the CFT central
charge can be attributed to quantum fluctuations in the geometry describing
the system. On the other hand, the logarithmic corrections arise from
thermal fluctuations and can be understood as quantum loop corrections Refs.
\cite{PO11,PO12,PO13}. Next, we focus our analysis on the bulk sound speed $%
c_{s}$ in transition is the ratio $c_{s}^{2}=\frac{\partial P}{\partial \rho
}=\frac{1}{\sqrt{C_{P}}}$ with $C_{p}=T\left( \frac{\partial S}{\partial T}%
\right) _{P}$ Ref. \cite{PO1,PO7}. Considering now $\rho =\frac{dM}{dV}$ and
$VdP=-PdV$ , we can easily check that%
\begin{equation}
\rho +P=T\frac{dS}{dV}+\mu _{d}\frac{dC}{dV}+\Phi \frac{dQ}{dV}+\Omega \frac{%
dJ}{dV}.  \label{b4}
\end{equation}%
Therefore from Eq. (\ref{a14}), the density behaves as%
\begin{equation}
\rho +\left( 1-\frac{1}{\chi }\right) P=T\frac{dS}{dV}+\Phi \frac{dQ}{dV}%
+\Omega \frac{dJ}{dV},  \label{b5}
\end{equation}%
yielding%
\begin{equation}
c_{s}^{2}=\frac{\partial P}{\partial \rho }=\frac{1}{\frac{1}{\chi }-1}.
\label{b6}
\end{equation}%
Such that the sound speed vanishes when $\chi =0$, which shows the
importance of the field $\chi $ which generates waves that propagate with
the speed $c_{s}$. Adiabatic compressibility is defined as $\kappa _{S}=-%
\frac{1}{V}\left( \frac{\partial V}{\partial P}\right) _{S,J,Q,...}=-\frac{1%
}{P}$, which is related to the speed of sound as $c_{s}^{-2}=1+\rho \kappa
_{S}$. It is interesting to notice that, in the above construction, $c_{s}$
played an essential role in the study of system stability. In order to
ensure $c_{s}^{2}>0$ we need $0<\chi <1$, which is in agreement with the
result of perturbative stability of Reissner-Nordstr\"{o}m black hole Refs.
\cite{D4,D5,PO10}. In the case of \ $\chi =0$, the sound speed vanishes,
indicating that the sound is due to the AdS pressure. For $\chi <0$ or $\chi
>1$ namely, $c_{s}^{2}<0$, i.e. the system is unstable. In the limit $1/\chi
\rightarrow 0$, we can use Taylor expand: $c_{s}^{2}=-1-\frac{1}{\chi }$.
Then it is clear that the specific heat reads as
\begin{equation}
C_{P}=\left( \frac{1}{\chi }-1\right) ^{2}>0.  \label{b7}
\end{equation}%
According to this last equation we see that the specific heat is strictly
positive Fig. (\ref{F1}), which signifies the thermodynamic stability of the
black hole.
Note that the divergent specific heat from $C_{P}<0$ to $C_{P}>0$ implies
the existence of a second-order phase transition. It is obviously seen that
when $\chi =0$, the specific heat vanishes. The parameter $\chi $ plays an
important role in describing black hole stability. For low $\chi $, as shown
in Fig. (\ref{F1}), that the $C_{P}$ for the first passage time decreases
quickly while $c_{s}$ increases. This means that evolution occurs in a short
time. The point $\chi =0.5$\ corresponds to a zero value of the adiabatic
compressibility $\kappa _{S}$. Also, at this point, we have $C_{P}=c_{s}=1$.
The point $\chi =1$\ corresponds to a zero heat capacity, which means that
an infinitesimally small amount of energy will increase the temperature by
an infinitely large amount. From the heat capacity viewpoint; there is a
transition of the small-large black holes at $\chi =1$.

\section{AdS thermodynamics}

\begin{figure}[H]
\centering\includegraphics[width=11cm]{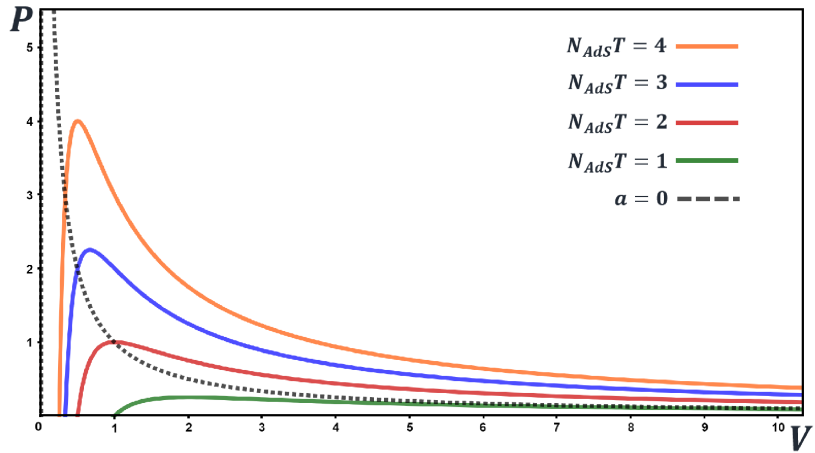}
\caption{Curves of some values $P-V$ isotherms of the van der Waals equation
of state ($N_{AdS}^{2}a_{t}=1$). The dashed line represents an ideal gas
(for $N_{AdS}T\ll 1$). While the solid line represents the real gas. In this
case, the ideal cases exist in the boundary of a real fluid.}
\label{F2}
\end{figure}

In order to obtain a duality between holographic CFT and the AdS
thermodynamics Ref. \cite{PO11,PO15}, we employ the following relation $%
dM\propto Cd\mu _{d}$. This suggests that the black hole first law reads
Ref. \cite{PO12}
\begin{equation}
dM=TdS+Cd\mu _{d}+VdP+\Phi dQ+\Omega dJ.  \label{c1}
\end{equation}%
In this case, $\mu _{d}=\left( \frac{\partial M}{\partial C}\right) $ is the
thermodynamic conjugate to the central charge. To find the holographic
entropy, we impose $S=\int \frac{Cd\mu _{d}}{T}$. From Eq. (\ref{a14}) we
get an entanglement entropy including the the boundary entropy: $S=C\frac{%
\mu _{d}}{T}\ln \mu _{d}\propto C$ Ref. \cite{RE6}. In this case, the
density behaves as%
\begin{equation}
dM=TdS+\Phi dQ+VdP+\frac{1}{2}\left( \mu _{d}dC+Cd\mu _{d}\right) .
\label{c2}
\end{equation}%
The corresponding entropy is then given by $S=\int \frac{1}{T}d\left( M+%
\frac{1}{2}\mu _{d}C\right) +S_{0}$, yielding $S=\int \frac{d\tilde{M}}{T}%
+S_{0}$ where
\begin{equation}
\tilde{M}=M+\frac{1}{2}\mu _{d}C.  \label{c3}
\end{equation}%
For a given mass $M$ and $\mu _{d}$, the mass $\tilde{M}$ is generated by
the presence of $C$. The parameter $\tilde{M}$ is related to the black hole
mass. Using the Euler relation $M=TS+\mu N$, it is obviously seen that $%
C\sim 2N$. Also when $\mu _{d}\sim N^{1-k}$ we get $C\sim N^{k}$. For
example, $C\sim N^{2}$, which corresponds to the result found by the
conformal symmetry in $SU(N)$ Ref. \cite{V0}. The number of degrees of
freedom simply as $C$ for high-energy states in large-N theories. Before we
proceed further, let us consider that $M=TS+\mu N_{AdS}$, which means that $%
C\sim 2N_{AdS}$. This leads to the following relation $C\sim \frac{\partial
N_{AdS}^{2}}{\partial N_{AdS}}.$ We note that the central charge$\ $is
invariant under the transformation $N_{AdS}^{2}\rightarrow N_{AdS}^{2}+K$,
where $K$\ is a constant. Then it is clear that $N_{CFT}^{2}\sim \frac{%
\partial }{\partial N_{AdS}}\left( N_{AdS}^{2}+K\right) .$ Let us consider
Eq. (\ref{a8}) $N_{CFT}^{2}\sim 2N_{AdS}$, this leads to express $G$ in the
following term $G\sim A^{2}/2L^{2}$, which shows that $G$\ can change with
scale, this may be the origin of the constant found in the rotation of
galaxies Ref. \cite{RE9}. Now, we would like to rewrite Eq. (\ref{c3}) in
terms of parameter $\chi $:
\begin{equation}
\tilde{M}=M+\frac{1}{2\chi }PV.  \label{c4}
\end{equation}%
From Ref. \cite{RE6} we have $M=E+PV$, where $E$ is the internal energy. In
the holographic frame, the parameter $\tilde{M}$ can be interpreted as a
gravitational version of chemical enthalpy Ref. \cite{RE8}. The total energy
of a system includes both its internal energy $\tilde{E}=M$ and the energy $%
PV$ required to displace the vacuum energy of its environment. It is
reasonable to assume that Eq. (\ref{c4}) is an extension of $M=E+PV$ in AdS.
Then it is clear that $\tilde{M}$ (in AdS) is the dual of $M$ (in CFT). We
can show the Van der Waals equation of the AdS fluid Ref. \cite{PO8}:%
\begin{equation}
P=\frac{N_{AdS}T}{V}-\frac{N_{AdS}^{2}a_{t}}{V^{2}}.  \label{c5}
\end{equation}%
where $T\equiv 2\chi \tilde{M}/N_{AdS}$, $a_{t}\equiv 2\chi VM/N_{AdS}^{2}$
is a measure of the average attraction between particles and $N_{AdS}$ is
the number of states associated with the horizon. When $N_{AdS}%
\longrightarrow \infty $, the attraction between particles vanishes $a_{t}=0$%
. Note that for $a_{t}=0$ we obtain the ideal gas law. When $T>T_{c}$, the
fluid is only stable under one phase: the supercritical fluid. While for $%
T<T_{c}$, the fluid is stable under a single phase, or present
simultaneously in two phases in equilibrium Ref. \cite{PO13}. We plot the
thermodynamic pressure $P$ as a function of the black hole volume with fixed
temperature $T$ and various parametrically in Fig. (\ref{F2}).
\begin{figure}[H]
\centering\includegraphics[width=11cm]{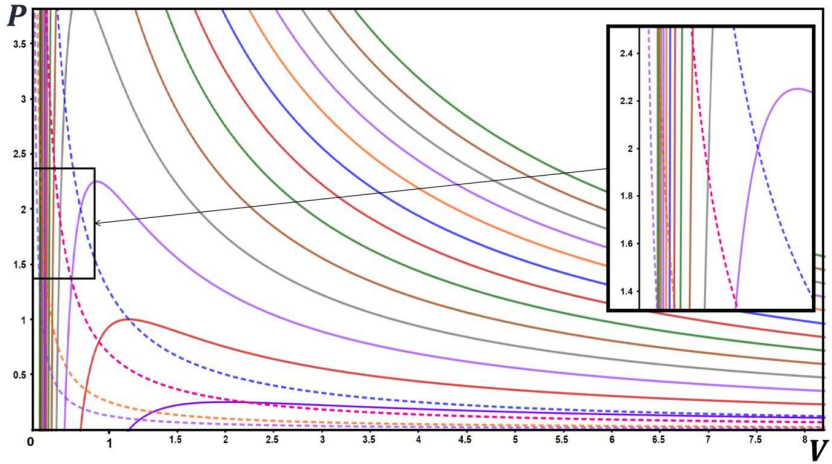}
\caption{Curves of some values $P-V$ isotherms graph represents the
behaviour of the ideal gas (dashed lines) and the real gas (solid lines).}
\label{F3}
\end{figure}
\begin{figure}[H]
\centering\includegraphics[width=9cm]{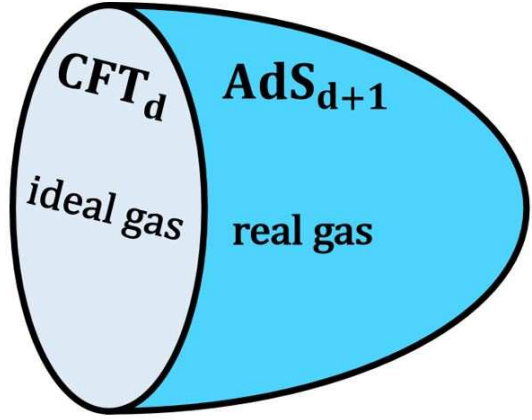}
\caption{Real/ideal gas duality in AdS/CFT correspondance.}
\label{F4}
\end{figure}
We notice the appearance of parameter $\chi $ in all the equations. If Eq. (%
\ref{a14}) describes an ideal gas in CFT and Eq. (\ref{c5}) describes a real
gas in AdS, this shows that the real gas is the dual of the ideal gas
according to the AdS/CFT correspondence Ref. \cite{IN0}. The average
attraction is written as
\begin{equation}
a_{t}=\frac{2\mu _{d}M}{N_{AdS}^{2}P}\chi ^{2}C.  \label{c6}
\end{equation}%
The attraction between the particles depends on the CFT central charge,
which shows that the black hole creates an interaction field for the
particles. For low pressure $P$ and low volume $V$, as shown in Fig. (\ref%
{F2}), the equation of state $P-V$ of an ideal gas (dashed line) creates a
boundary for the behavior of a real gas (solid lines). In Fig. (\ref{F3}),
we notice that the region at a low volume behaves like a real and ideal gas
horizon. For each ideal gas curve, there is a surface of the real gas
curves, which shows that the ideal gas is the boundary of the real gas. Then
it is clear that there is a great similarity between the AdS/CFT duality and
the duality between real and ideal gas Fig. (\ref{F4}). Next, we study the
relation between the real and ideal gas by the AdS/CFT correspondence. From
Eq. (\ref{a14}) and $Z_{CFT}=\exp (-\frac{\mu C}{T})$ we have%
\begin{equation}
\ln Z_{CFT}=-\frac{\mu }{\mu _{d}\chi }\frac{PV}{T}.  \label{c7}
\end{equation}%
Using the AdS/CFT correspondence ($Z_{AdS}=Z_{CFT}$), the ideal gas equation
$PV=N_{CFT}T$ Ref. \cite{PO5} and the real gas ( with $T_{AdS}=T_{\func{real}%
})$ equation (\ref{c5}), so we find
\begin{equation}
\exp (-\frac{\mu }{\mu _{d+1}\chi }\frac{PV}{T})_{\func{real}}=\exp (-\frac{%
\mu }{\mu _{d}\chi }N_{CFT})_{\text{ideal}}.  \label{c8}
\end{equation}%
The corresponding temperature is $T_{\func{real}}=\frac{2\mu _{d}}{%
N_{CFT}\mu _{d+1}}\left( \tilde{M}-M\right) \chi $ with $\mu _{d+1}/\mu _{d}$
is a constant. This shows that the AdS temperature will be zero if $M=\tilde{%
M}$. i.e. $\tilde{M}$ is a critical mass that describes an AdS space-time
without temperature ($Z_{AdS}=0$). We notice that when the temperature
increases, the black hole mass $M$ decreases. Also if $N_{CFT}\rightarrow 0$%
, the temperature of AdS increases. The AdS equation of state is given by%
\begin{equation}
\left. PV\right\vert _{AdS}=\frac{\mu _{d+1}}{\mu _{d}}N_{CFT}T_{\func{real}%
}.  \label{c9}
\end{equation}%
According to this equation, when the real gas moves from AdS to CFT, it
behaves like an ideal gas. The presence of the ($\mu _{d},$ $\mu _{d+1}$)
exhibits the AdS/CFT duality for this black hole Ref. \cite{MB1}. This shows
that the thermodynamic parameters of AdS are expressed as a function of the
physical parameters of CFT.

\section{Extended thermodynamics of 4d AdS-Kerr black hole}

In order to discuss the applied real/ideal gas duality, it is pertinent to
study for example the case of the 4-dimensional Kerr-AdS spacetime Ref. \cite%
{PO2,PO9}. The Kerr-AdS metric describes a rotating spacetime by the
spherical coordinates $\left( t,r,\theta ,\varphi \right) $ as%
\begin{equation}
ds^{2}=-\frac{\Delta }{\rho ^{2}}\left( dt-\frac{a\sin ^{2}\theta }{\Xi }%
d\varphi \right) ^{2}+\frac{\Sigma \sin ^{2}\theta }{\rho ^{2}}\left[ \frac{%
r^{2}+a^{2}}{\Xi }d\varphi -adt\right] ^{2}+\frac{\rho ^{2}}{\Delta }dr^{2}+%
\frac{\rho ^{2}}{\Sigma }d\theta ^{2},  \label{d1}
\end{equation}%
with $\Delta \equiv \left( r^{2}+a^{2}\right) \left( 1-\frac{a^{2}}{L^{2}}%
\right) -2GMr$, $\rho ^{2}=r^{2}+a^{2}\cos ^{2}\theta $, $\Xi =1-\frac{a^{2}%
}{L^{2}}$, $\Sigma =1-\frac{a^{2}}{L^{2}}\cos ^{2}\theta $ and $J=aM$ is the
angular momentum. The Kerr metric describes a black hole if and only if $%
a\leq M^{2}$ or $M^{2}\geq a^{2}$. Then it is clear that the horizons in a
Kerr spacetime are located at $r_{\pm }=M\pm \sqrt{M^{2}-a^{2}}$ with $%
r_{+}=r_{H}$ Ref. \cite{PO1}. The outer and inner event horizons are located
at $\Delta (r_{\pm })=0$ with $T_{\pm }=\frac{\Delta ^{\prime }\left( r_{\pm
}\right) }{4\pi }$. For Kerr-AdS black holes the left and right CFTs have
identical central charges are%
\begin{equation}
C=C_{L}=C_{R}=\frac{6a\left( r_{+}^{2}-r_{-}^{2}\right) }{\Delta ^{\prime
}\left( r_{+}\right) },  \label{d2}
\end{equation}%
For the Kerr-AdS black hole, the Hawking temperature and the angular
velocity are%
\begin{equation}
T=\frac{r_{+}}{4\pi \left( r_{+}^{2}+a^{2}\right) }\left( 1+\frac{a^{2}}{%
L^{2}}+3\frac{r_{+}^{2}}{L^{2}}-\frac{a^{2}}{r_{+}^{2}}\right) .
\end{equation}%
\begin{equation}
\Omega =a\frac{1+\frac{r_{+}^{2}}{L^{2}}}{r_{+}^{2}+a^{2}}.
\end{equation}%
Note that for the ideal gas given implicitly by Eq. (\ref{a14}), the
critical central charge, corresponds to
\begin{equation}
\left. \frac{\partial T}{\partial r_{+}}\right\vert
_{r_{H}=r_{c},C=C_{c}}=\left. \frac{\partial ^{2}T}{\partial r_{+}^{2}}%
\right\vert _{r_{H}=r_{c},C=C_{c}}=0.  \label{d8}
\end{equation}%
The Bekenstein-Hawking entropy is\newline
\begin{equation}
S=\frac{\pi \left( r_{+}^{2}+a^{2}\right) }{\Xi G}+\frac{1}{\chi }\ln \frac{%
C_{0}}{C}.
\end{equation}%
Here we actually assume that the left and right central charges should be
the same$.$ From Eqs. (\ref{a14},\ref{d2}), we derive the Kerr-AdS equation
of the state
\begin{equation}
PV=\frac{6a\mu _{d}\left( r_{+}^{2}-r_{-}^{2}\right) }{\Delta ^{\prime
}\left( r_{+}\right) }\chi .  \label{d3}
\end{equation}%
Then it is clear that%
\begin{equation}
\tilde{M}=M+\frac{3a\mu _{d}\left( r_{+}^{2}-r_{-}^{2}\right) }{r_{+}\left(
1-\frac{GM}{r_{+}}-\frac{a^{2}}{L^{2}}\right) }.  \label{d4}
\end{equation}%
where $\Delta ^{\prime }\left( r_{+}\right) \equiv 2r_{+}\left( 1-\frac{GM}{%
r_{+}}-\frac{a^{2}}{L^{2}}\right) $. The Cardy entropy formula for the
AdS-Kerr black holes is $S_{\pm }=\frac{\pi ^{2}}{3}\left( C_{L}T_{L}\pm
C_{R}T_{R}\right) $. The identical central charges $C_{L}=C_{R}$ provide
evidence for the macroscopic description of the black hole by the Cardy
formula. The left and right CFT number $N_{\pm }$ of microscopic degrees of
freedom for the AdS-Kerr black hole is
\begin{equation}
N_{AdS}=\frac{L^{2}}{G}\text{, \ }N_{CFT}^{\pm }=\frac{4\pi r_{\pm }^{2}}{G}%
\text{,}  \label{d5}
\end{equation}%
Let us mention that for AdS-Kerr black hole, the AdS states are distributed
along bulk, while, the CFT states degenerate on both horizons.%
\begin{equation}
N_{CFT}^{+}-N_{CFT}^{-}=\frac{4\pi r_{+}}{3Ga\mu _{d}}\left( 1-\frac{GM}{%
r_{+}}-\frac{a^{2}}{L^{2}}\right) \left( \tilde{M}-M\right) .  \label{d6}
\end{equation}%
Note that for \textquotedblleft small\textquotedblright\ Kerr-AdS black
holes we have $N_{CFT}^{+}\approx N_{CFT}^{-}$. This expression gives an
interpretation of the difference $\tilde{M}-M$, which describes the mass gap
between the two horizons of the Kerr black hole. This shows the difference
between the AdS (real) gas and the holographic (ideal) gas of CFT. For the
small-large black hole phase transition Ref. \cite{RE10} below the critical
point, the change of the horizon radius difference $\tilde{M}-M$ becomes
zero at the critical point. From Eq. (\ref{d6}) we get $%
N_{CFT}^{+}-N_{CFT}^{-}=\frac{4\pi }{3a\mu _{d}}\left( \hat{M}-M\right)
\left( \tilde{M}-M\right) $, where $\Xi =\frac{G\hat{M}}{r_{+}}=1-\frac{a^{2}%
}{L^{2}}$ is a gravitational potential is related to the AdS geometry, while
$\frac{G\tilde{M}}{r_{+}}$ is the gravitational potential that is generated
by both the mass of the black hole and its CFT central charge. In AdS-Kerr
space, we introduce a new gravitational potential field%
\begin{equation}
\Phi \left( r\right) \equiv \frac{\Delta }{r^{2}}=\left( 1+\frac{a^{2}}{r^{2}%
}\right) \left( 1-\frac{a^{2}}{L^{2}}\right) -\frac{2GM}{r}.  \label{d7}
\end{equation}%
We note that the above equation$\ $is invariant under the transformation $%
r^{2}\longleftrightarrow -L^{2}$. Then it is clear that for $a^{2}=L^{2}$,
the potential $\Phi \left( r\right) $ will be that of the Schwarzschild
black hole. Thermodynamic quantities for the AdS-Kerr black hole may be
calculated by setting the factors $\chi $. To facilitate this we introduce a
new base
\begin{equation}
\chi _{i}=\left\{ \chi _{L}=\frac{r_{+}^{2}}{L^{2}},\chi _{a}=\frac{a^{2}}{%
L^{2}}\right\} ,\text{\ }\frac{a^{2}}{r_{+}^{2}}=\frac{L^{2}}{r_{+}^{2}}%
\frac{a^{2}}{L^{2}}=\frac{\chi _{a}}{\chi _{L}}.
\end{equation}%
In order to obtain general transformation in 4 dimensions we need to rewrite
the thermodynamic parameters of the AdS-Kerr back hole otherwise:%
\begin{equation}
\Phi \left( r_{+}\right) =\left( 1+\frac{\chi _{a}}{\chi _{L}}\right) \left(
1-\chi _{a}\right) -\frac{2GM}{r_{+}}.
\end{equation}%
\begin{equation}
S=\left[ \frac{1+\frac{\chi _{a}}{\chi _{L}}}{1-\frac{a^{2}}{L^{2}}}\right]
\frac{\pi L^{2}}{G}\chi _{L}+\frac{1}{\chi _{L}^{2}}\ln \frac{C_{0}}{C}.
\end{equation}%
\begin{equation}
\Omega =\left[ \frac{1+\chi _{L}}{1+\frac{\chi _{a}}{\chi _{L}}}\right]
\frac{a}{L^{2}\chi _{L}}.
\end{equation}%
\begin{equation}
T=\left[ \frac{1+\chi _{a}+3\chi _{L}-\frac{\chi _{a}}{\chi _{L}}}{1+\frac{%
\chi _{a}}{\chi _{L}}}\right] \frac{1}{4\pi r_{+}}.
\end{equation}%
This indicates that the parameter $\chi _{i}$ plays a very important role in
the thermodynamics of the AdS-Kerr black hole. Next, we use the impact
parameter $b=\left\vert \hat{L}\right\vert /E$ in 4-dimensional space-time.
We easily get the equation of radial motion $V_{eff}+\dot{r}^{2}=1/b^{2}$.
The effective potential for a free photon can be expressed as%
\begin{equation}
V_{eff}=\frac{\hat{L}^{2}}{r^{2}}f(r)-E^{2},
\end{equation}%
To study the photon orbit, we set $\Delta =d\Delta /dr=0$. The effective
potential for a massless particle is given by Ref. \cite{RK2}
\begin{equation}
V_{eff}\left( r,b\right) =-\frac{\Xi ^{2}}{b^{2}r^{4}}\left[ \left(
r^{2}+a^{2}-ab\right) ^{2}-\left( a-b\right) ^{2}\Delta \right] +\frac{1}{%
b^{2}},
\end{equation}%
which read%
\begin{equation}
V_{eff}\left( r,b\right) =\left( 1-\frac{a}{b}\right) ^{2}\frac{\Xi ^{2}}{%
r^{2}}\Phi \left( r\right) +\frac{1}{b^{2}}\left( 1-\frac{\Xi ^{2}}{r^{4}}%
\left( r^{2}+a^{2}-ab\right) ^{2}\right) .
\end{equation}%
For $a=b$ and $\Xi ^{2}\ll r^{4}$ we obtain $V_{eff}\approx \frac{1}{b^{2}}$%
, which corresponds to the result found in the context of the shadow of 4d
Einstein-Gauss-Bonnet black holes Ref. \cite{w5}.

\section{Conclusion}

In this paper, we investigated the holographic and thermodynamic aspects of
black holes in the context of the gauge/gravity duality. Motivated by the
AdS/CFT correspondence, we have expressed all the thermodynamic parameters
of the black hole according to a new parameterization. \newline
In the present work we have derived the modified Friedmann equation from an
d-dimensional Smarr formula, for perfect fluid, by assuming Barrow entropy
for the horizon. We explored how this parameterization affects the influence
on photon orbit and $PV$ energy. Our novel results demonstrate that the AdS
radius is the critical shadow radius. We have shown that the CFT particles
form an ideal fluid, while the AdS fluid forms a real gas and undergoes the
Van der Waals equation. From a phenomenological point of view, the ideal gas
represents the boundary of a real gas. We have studied the central charge of
the d-dimensional CFT. We have obtained the Hawking-Bekenstein formula with
logarithmic corrections, which depends on the central charge. We have
demonstrated that gauge/gravity duality (for gas) provides a new
understanding of the phase behavior of charged AdS black holes. We find that
phase behavior is governed by a critical value of the central charge instead
of the pressure. We have studied the dynamical process of the small/large
black hole phase transition governed by the heat capacity. First, we
reviewed the thermodynamics for the AdS-Kerr black hole. Second, we have
shown the difference in the number of CFT and AdS particles, which is caused
by the difference between the AdS-Kerr black hole mass and the bulk mass.

\end{document}